# Simulating Organogenesis in COMSOL

Philipp Germann[1], Dzianis Menshykau[1], Simon Tanaka[1] and Dagmar Iber[*,1,2]
[1]Departement for Biosystems Science and Engineering, ETH Zurich
[2]Swiss Institute of Bioinformatics (SIB), Switzerland
*Corresponding author: Mattenstrasse 26, CH-4058 Basel, dagmar.iber@bsse.ethz.ch

**Abstract:** Organogenesis is a tightly regulated process that has been studied experimentally for decades. Computational models can help to integrate available knowledge and to better understand the underlying regulatory logic. We are currently studying mechanistic models for the development of limbs, lungs, kidneys, and bone. We have tested a number of alternative methods to solve our spatio-temporal differential equation models of reaction-diffusion type on growing domains of realistic shape, among them finite elements in COMSOL Multiphysics. Given the large number of variables (up to fifteen), the sharp domain boundaries, the travelling wave character of some solutions, and the stiffness of the reactions we are facing numerous numerical challenges. To test new ideas efficiently we have developed a strategy to optimize simulation times in COMSOL.

## 1. Introduction: Mechanistic Models for Organogenesis

Organogenesis is the process by which stem cells develop into organs in animals. In several systems important genes have been identified and the regulatory logic has been analyzed extensively over the last decades. The discovered regulatory networks are too complex to be understood intuitively and many questions remain open.

Organogenesis is a tightly regulated process, e.g. the lungs of two genetically identical embryos branch the same way [1]. This allows for deterministic modelling, which has been applied for decades to describe pattern formation in developmental biology [2]. Following this approach we were able to predict novel genetic regulations in the limb bud [3] and suggest a mechanism for lung branch mode selection (Menshykau et al., submitted) based on models implemented in COMSOL multiphysics, which has previously been shown to solve similar problems with a known analytic solution accurately [5].

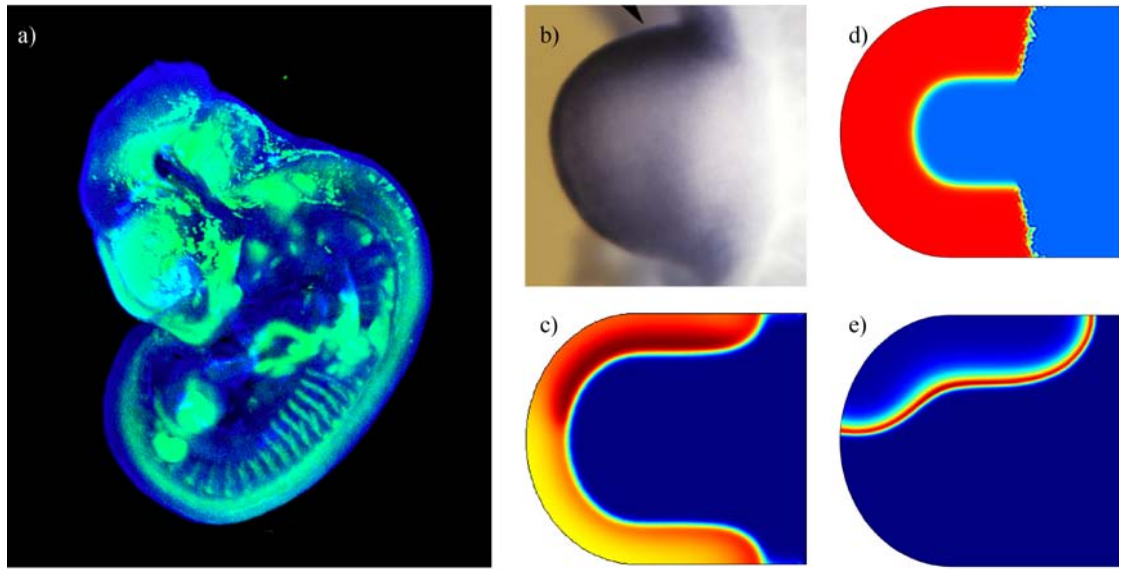

**Figure 1.** Modelling gene expression patterns in the mouse limb bud. (a) Optical Projection Tomography (OPT) provides 3D information on gene expression patterns in the mouse limb bud (image courtesy of Frédéric Laurent and Emanuele Pignatti). (b) Classical in situ staining provides a 2D projection of the expression data. The image is a reproduction of BMP4 expression data from Fig 6B in [4]. (c) The experimental data can then be compared to the predicted spatial distributions of expression rates. (d,e) For some variables our model produces sharp domain boundaries (d) and traveling waves (e).

Our models are formulated as systems of reaction-diffusion equations of the form

$$\dot{X}_i + \nabla(u \cdot X_i) = D_i \nabla^2 X_i + R_i$$

where $u$ denotes the velocity of the domain and $R_i$ the reactions, which couple the equations for the different species $X_i$. $D_i$ is the diffusion constant and $\nabla$ the Nabla operator. The velocity might be imposed or based on concentrations of proteins, which change properties of the cells, like division rate or adhesion.

Our models typically involve three to fifteen species and typical reactions describe decay

$$R_X = -\delta \cdot X$$

and complex formation

$$R_X = -k^+ \cdot m \cdot X^m \cdot Y^n + k^- \cdot m \cdot X_m Y_n$$

$$R_Y = -k^+ \cdot n \cdot X^m \cdot Y^n + k^- \cdot n \cdot X_m Y_n$$

$$R_{X_m Y_n} = k^+ \cdot X^m \cdot Y^n - k^- \cdot X_m Y_n$$

where $X_m Y_n$ stands for the complex made of $m$ $X$ and $n$ $Y$ molecules. The reaction terms can contain also other non-linear functions like enzymatic activation $\sigma$ and inhibition $\bar{\sigma} = 1 - \sigma$, where $\sigma$ is modelled analogous to Hill kinetics

$$\sigma = X^n / (X^n + K^n).$$

The threshold $K$ is the concentration at which the activation reaches half its strength and the exponent $n$ depends on the cooperativity of the regulating interactions. For example

$$R_X = \rho \cdot \sigma(Y)$$

describes a production term for a protein $X$ induced by another protein $Y$.

## 2. Advances: Optimizing COMSOL Models

The non-linearities and different timescales in the reaction and diffusion terms produce traveling wave and sharp edges in the solutions, potentially rendering the models hard to solve numerically. In addition these equations have to be coupled to equations from continuum mechanics describing growth. Since very little is known about the parameter values and even the interactions are far from carved in stone we need efficient tools to explore multiple possibilities and together with the COMSOL support we have developed a strategy to optimize models in COMSOL.

Even before tuning solver settings singularities can be reduced by smoothening sharp corners with the fillet node and by using the inbuilt smooth step function, e.g. in initial conditions or spatially restricted reactions.

Some logarithmic derivatives in our models are very large in isolated points. The values of the different concentrations vary over several orders of magnitude and mask this stiffness. This leads the solver to take too large timesteps resulting in divergences in complex formations. Hence the next step is to produce a complete solution at any cost in order to identify the variables causing these spikes using a 'sledge-hammer method'. This might include limiting the timesteps to very small values, updating the Jacobian at each iteration, tolerating only very small relative and absolute errors and using a very fine mesh. The linear solver MUMPS was found to be the most stable in our tests.

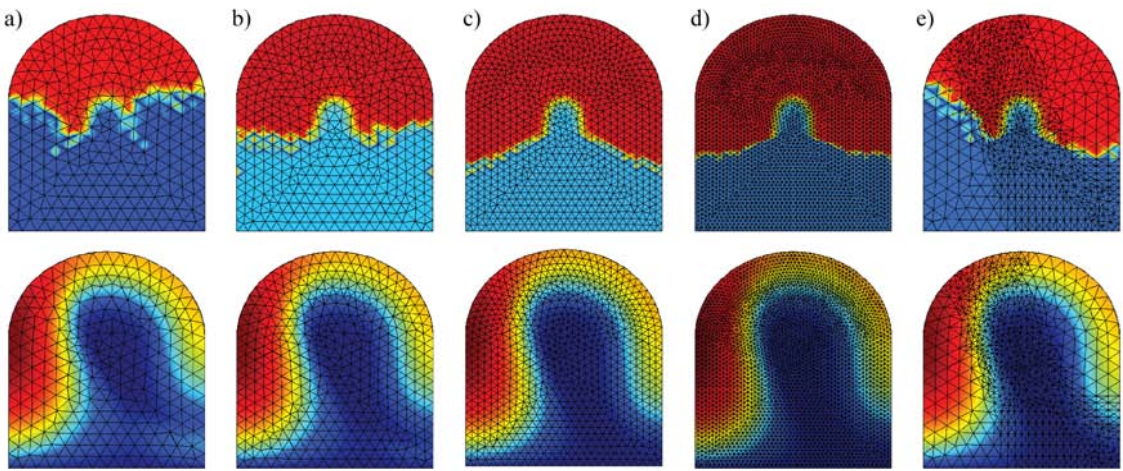

**Figure 2**. Limb bud patterning: the impact of meshing on two of fifteen species. The first column (a) shows 528 elements and (b) 865 elements, both simulations take roughly half an hour on four cores. The calculation in (c) with 1749 elements runs 20% faster and (d) with 5847 elements takes three and a half hours. In the last column (e) artificial asymmetry from adaptive remeshing is shown on a calculation with roughly 1700 elements. The mesh is remeshed according to the gradient of the traveling wave of complex formation introduced in figure 1 (e).

After scaling the variables in COMSOLs solver configurations node against the maximal values obtained we can relax the solver settings back to automatic time stepping and automatic dampening (updating the Jacobian). These optimized simulations also run on a very coarse mesh, but not in all cases fast and accurate, c.f. figure 2 (a) to (d). The linear solver PARDISO turned out to be the fastest.

Using this strategy we were able to reduce computing times from initial 35 hours to 25 minutes in the limb bud model depicted in figures 1 and 2.

The oscillations around the sharp edges persist, but converge upon refining the mesh. Using COMSOLs consistent stabilization feature available in the chemical species transport module prevents these oscillations, but prolongs calculation times and might impact the resulting patterns.

Furthermore variables can be grouped and different solver settings applied sequentially to these groups using COMSOLs segregated solver feature. Random segregation in two or three groups quadruplicated the computing time; on the other hand segregating variables based on biological considerations (e.g. co-regulation of patterning events such as proximal-distal and anterior-posterior axis formation) allowed us to reduce the computing time below nine minutes. Interestingly isolating the variables that together caused spikes in our initial model allowed us to reduce the computing time even further from 25 minutes to five minutes.

COMSOL was reported to provide significant speed-up for a problem with 1.3 million degrees of freedom [6]. We also ran our limb bud model in parallel on a single node (i.e. no MPI), but the speed-up turned out to be poor. While our limb bud model may have been too small for efficient parallelization, further tests with larger models in three dimensions did not yield any improvements (figure 3). The data was collected using PARDISO, but the other linear solvers did not perform better.

We also benchmarked adaptive remeshing since our test model exhibits localized features, like the traveling waves and sharp edges. However it did not accelerate the calculations further and adaptive remeshing can introduce artificial asymmetries as shown in figure 2 (e), since for instance the effective diffusion depends on the discretization.

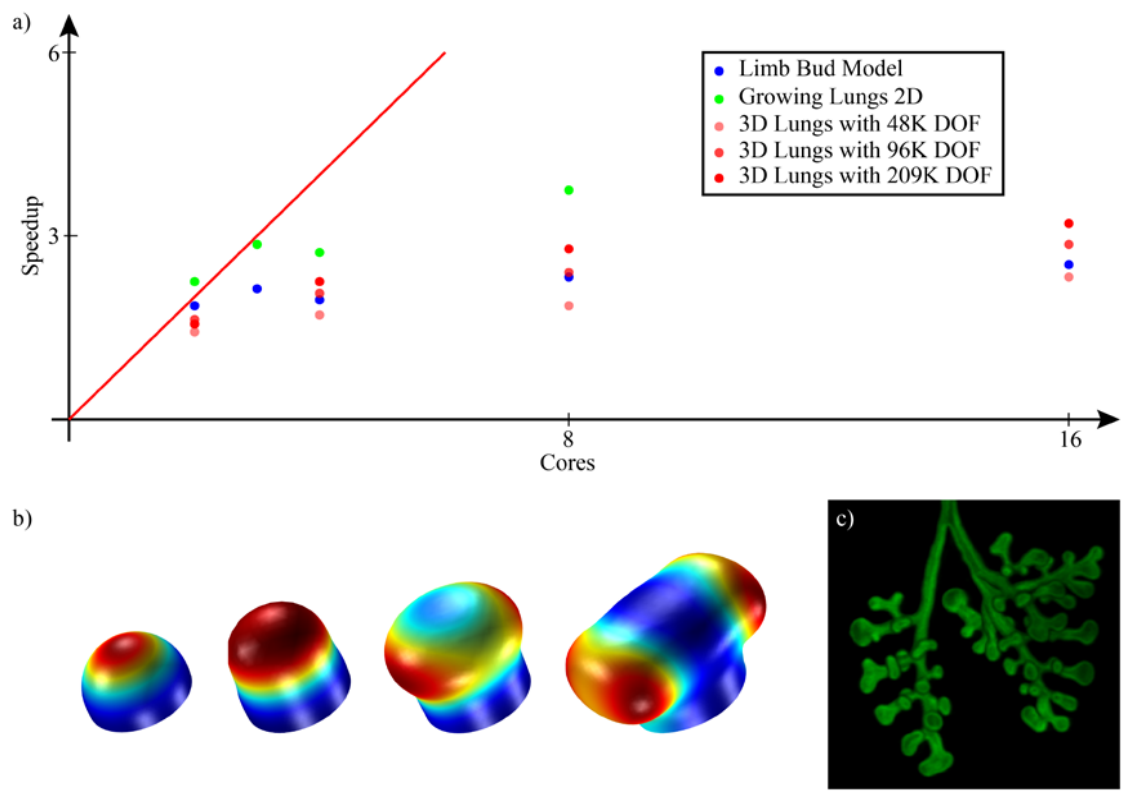

**Figure 3.** (a) Speed-up of different models upon parallelization (strong scaling). The red line shows the theoretically maximal possible linear speed-up and the dots are ratios of computing time at a certain number of cores divided by computing time using a single core. (b) The 3D lung branching model referred to in (a). Its shape arises from displacements along the surface normal at velocities proportional to the morphogen concentration shown on the surface. (c) Three dimensional imaging data of the lung epithelium in a developing mouse embryo.

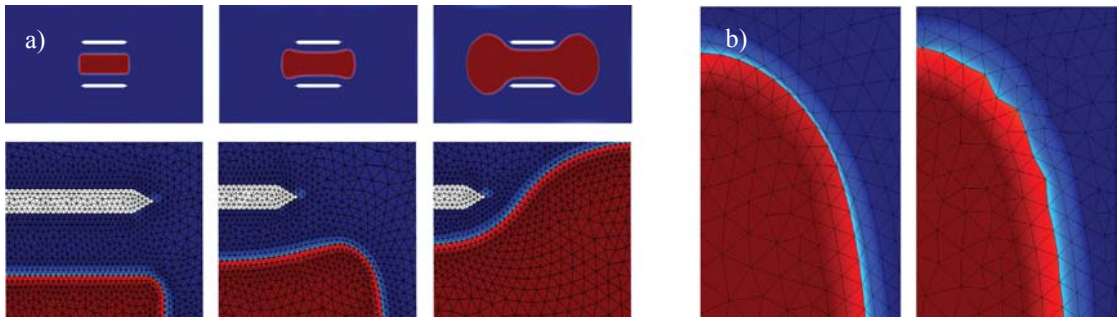

**Figure 4.** The new remeshing feature illustrated with a model for long bone growth (a). It consists of a fluid governed by the Navier-Stokes equations, with a mass source as plotted, between two walls, simulated using the creeping flow node. The mesh quality was kept above 0.4 with 42 remeshing steps. On the right side a comparison between Laplacian and Winslow smoothing is given (b).

### 3. Challenges: Large Deformations

To describe deforming domains due to growth we applied the ALE moving mesh module. Since the meshes of the subdomains adjacent to the moving boundary get distorted and stretched or squeezed, the mesh quality quickly worsens. The recent COMSOL version 4.2 features automatic remeshing to overcome this, cf. figure 4 (a). The principle is simple: a measure for quality is calculated regularly for each mesh element. Whenever this mesh quality falls below a predefined barrier, the entire domain is remeshed. In spite of not yet fully exhausting the possibilities we give a short summary of the experiences collected so far.

For uniform models with smooth deformations automatic remeshing usually worked well, but when more sophisticated meshing settings, e.g. different properties for subdomains, are needed they might no longer be appropriate due to large deformations and the mesher may no longer be able to create a mesh with the desired quality leading to abortion.

Setting the shape order to linear reduces the number and delays the appearance of inverted elements and Laplacian smoothing worked best for avoiding inverted elements, cf. figure 4 (b).

The meshing and fine resolution of the moving boundary are crucial for successfully running the bone model depicted in figure 4 with large deformations.

When a highly resolved moving boundary moves too close to a low-resolution external boundary during a simulation restricting element growth can save the meshing algorithm from failing.

Solver settings similar to those described as 'sledge hammer method' in the previous section allowed simulations to run further. Enforcing frequent remeshing by demanding high mesh quality further supported this.

### 4. Conclusions & Outlook: Studying Growth in 3D

COMSOLs powerful interface and vast features allow us to implement new ideas quickly and to test them efficiently. Based on benchmarks with Discontinuous Galerkin Methods implemented in DUNE-FEM [7] we expect that our computing times are in a reasonable range.

With the recent automatic remeshing feature of COMSOL it becomes technically feasible to run our models on realistically growing domains. This requires us to couple the gene regulatory networks to fluid or solid-state equations, which creates additional numerical difficulties. Preliminary simulations of the limb bud model shown in figures 1 and 2 in three dimensions required several days to run. In spite of important advances in our computational workflow there are still large challenges ahead.

### 5. References

1. Metzger, R. J., Klein, O. D., Martin, G. R. and Krasnow, M. A., The branching programme of mouse lung development, *Nature*, **453**, 745-750 (2008)

2. Kondo, S. and Miur, T., Reaction-diffusion model as a framework for understanding biological pattern formation. *Science,* **329**, 1616–1620 (2010)

3. Probst, S., Kraemer, C., Demougin, Ph., Sheth, R., Martin, G. R., Shiratori, H., Hamada, H., Iber, D., Zeller, R. and Zuniga, A., SHH propagates distal limb bud development by enhancing CYP26B1-mediated retinoic acid clearance via AER-FGF signalling, *Development*, **138**, 1913-23 (2011)

4. Galli, A., Robay, D., Osterwalder, M., Bao, X., Bénazet, J.-D., Tariq, M., Paro, R., Mackem, S., Zeller, R., Distinct Roles of Hand2 in Initiating Polarity and Posterior Shh Expression during the Onset of Mouse Limb Bud Development, *PLoS Genetics*, **6**, e1000901 (2010)

5. Thümmler, V. and Weddemann, A., Computation of Space-Time Patterns via ALE Methods, *Exerpt from the Proceedings of the COMSOL Users Conference 2007 Grenoble*

6. http://www.comsol.com/shared/downloads/partners/Comsol_datasheet_Final_Web.pdf

7. Dedner, A., Klöfkorn, R., Nolte, M. and Ohlberger, M., A Generic Interface for Parallel and Adaptive Scientific Computing: Abstraction Principles and the DUNE-FEM Module, *Computing*, **90**, 165-196 (2010)

## 6. Acknowledgements

We would like to thank the COMSOL support team and especially Sven Friedel for all the help, Robert Klöfkorn for the implementation of our benchmark problem with Discontinuous Galerkin Methods in DUNE and Rolf Zeller and his group with Frédéric Laurent and Emanuele Pignatti for the images and the ongoing collaboration. D. M. is funded by an ETH Zurich postdoctoral fellowship.